\documentclass[a4paper,11pt]{article}
\usepackage[utf8]{inputenc}
\usepackage[T1]{fontenc}
\usepackage{authblk}
\usepackage{amsmath,amssymb}
\usepackage{graphicx}
\usepackage[hmargin=2.5cm,vmargin=3cm]{geometry}
\usepackage{booktabs}
\usepackage{braket}
\usepackage{enumerate}
\usepackage{mathtools}
\usepackage{microtype}
\usepackage[numbers,sort&compress]{natbib}
\usepackage{tensor}
\usepackage{textcomp}
\usepackage[inline]{enumitem}
\usepackage{siunitx}
\usepackage[scr=rsfso]{mathalpha}
\RequirePackage[colorlinks=true
,urlcolor=blue
,citecolor=blue
,linkcolor=blue
]{hyperref}
\DeclareSIUnit\lightspeed{\text{\ensuremath{c}}}
\DeclareSIUnit\boltzmannk{\text{\ensuremath{k_{\textup{B}}}}}

\newcommand*\idd[1]{\mathrm{d}#1\,}
\newcommand*\iddn[2]{\mathrm{d}^{#1}#2\,}
\newcommand*{\order}[1]{\mathcal{O}(#1)}
\newcommand*{\pheq}{\mathrel{\phantom{=}}}

\DeclarePairedDelimiterX{\abs}[1]{\lvert}{\rvert}{#1}
\title{Matter--antimatter asymmetry and non-inertial effects}

\author[1]{V. M. G. Silveira\footnote{viniciusmgsilveira@gmail.com}}

\author[1,2]{C. A. Z. Vasconcellos\footnote{cesar.zen@ufrgs.br}}

\author[1]{E. G. S. Luna\footnote{luna@if.ufrgs.br}}
 
\author[1]{D. Hadjimichef\footnote{dimiter.hadjimichef@ufrgs.br}}

\affil[1]{Instituto de Física, Universidade Federal do Rio Grande do Sul, Caixa Postal 15051, Porto Alegre -- RS, 91501-970, Brazil}
\affil[2]{International Center for Relativistic Astrophysics Network (ICRANet), Pescara, Italy}

\begin{document}
\maketitle

\begin{abstract}
 We investigate non-inertial effects on \(CP\)-violating processes using a model, based on the framework of quantum field theory in curved spacetimes, devised to account for the decay of accelerated particles. We show that the \(CP\)~violation parameter for the decay of accelerated kaons into two pions decreases very slightly as very high accelerations are achieved, implying decreased asymmetry between matter and antimatter in this regime. We discuss the relationship between these results and cosmological processes surrounding matter--antimatter asymmetry and argue that, due to the connection between non-inertial and thermal phenomena established by the Unruh effect, this kind of computation may prove useful in furthering the understanding of thermodynamical effects in curved spacetimes.
\end{abstract}

%%%%%%%%%%%%%%%%%%%%%%%%%%%%%%%%%%%%%%%%%%%%%%%%%%%%%%%%%%%%%%%%%%%%%%%%%%%%%%%%%

\section{Introduction}

 One of the great modern scientific mysteries is the abundance of matter over antimatter. Cosmic matter--antimatter asymmetry arises in the framework of elementary particle physics through baryogenesis models, which offer mechanisms for obtaining matter--antimatter asymmetry from an initially symmetric Universe~\cite{Rubakov}. It is well established that baryogenesis requires three crucial ingredients~\cite{Sakharov}:
 \begin{enumerate*}[label=(\roman*)]
  \item baryon number violation,
  \item violation of \(C\) (particle--antiparticle) symmetry and the combination~\(CP\) of \(C\) and \(P\) (left--right or parity) symmetries and
  \item departure from thermal equilibrium.
 \end{enumerate*}

 In the Standard Model~(SM) the matter--antimatter asymmetry of the Universe is credited to \emph{\(CP\) violation}~(\(CP\)v), although experimental evidence has systematically shown that it may not be sufficient to explain this imbalance~\cite{Farrar,Huet}. \(CP\)v has been observed in various weak decays involving strange and beauty quarks, being recently confirmed by the LHCb collaboration for the charmed \(D\)~meson~\cite{LHCb}. The main sources of \(CP\)v in the SM are 
 \begin{enumerate*}[label=(\roman*)]
  \item the quark sector, involving the Cabibbo--Kobayashi--Maskawa (CKM) matrix,
  \item the strong interaction and,
  \item the Pontecorvo--Maki--Nakagawa--Sakata (PMNS) matrix in the lepton sector \cite{PDG}.
 \end{enumerate*}
 The CKM~matrix has been observed experimentally and can only account for a small portion of the \(CP\)v required to explain the matter--antimatter asymmetry. The failure, until now, to observe the electric dipole moment of the neutron in experiments suggests that any \(CP\)v in the strong sector is also too small to account for the necessary matter--antimatter asymmetry in the early Universe. In the case of neutrinos being Majorana fermions, for example, the PMNS matrix could have two additional \(CP\)-violating Majorana phases, which would lead to a new source of \(CP\)v in the framework of the SM in the lepton sector. Alternatively, \(CP\)v in the lepton sector could result, experimentally, to be too small to account for matter--antimatter asymmetry, but additional sources of \(CP\)v could arise from some still unknown physics beyond the SM.

 In general, \(CP\)-violating observables are computed from decay amplitudes, decay rates, or quantities derived from these, such as a particle's lifetime, which is obtained from the decay rates and is regarded as one of its inherent and characteristic properties. It is well known since the seminal work of Fulling, Davies and Unruh~\cite{Fulling,Davies,Unruh} in quantum field theory in curved spacetimes~(QFTCS) that a uniformly accelerated detector moving through the usual flat spacetime vacuum of a conventional quantum field theory responds as though it were in a thermal bath of temperature
 \begin{equation}
  T = \frac{\hslash a}{2\pi k_{\textup{B}}c},
 \end{equation}
 where \(a\)~is the acceleration (this is often referred to as the \emph{Unruh effect}). One may, thus, expect that the acceleration would causes a modification of particle lifetimes with respect to their own proper time, i.e., in their accelerated rest frame, much as a thermal bath would do to a particle in an inertial frame. This is indeed the case---as was shown by Müller~\cite{Muller} and Vanzella and Matsas~\cite{VanzellaShort,VanzellaLong} in their investigation of the decay rates of accelerated particles---and the objective of this paper is to investigate the impact these non-inertial effects on \(CP\)-violating observables, i.e., we are interested in the non-inertial factor of the following heuristic equation,
 \begin{equation}
  CP\text{v}= (\text{imbalance}) \times (\text{non-inertial modification}).
 \end{equation}
 Specifically, we investigate these effects on the decay rates and in a \(CP\)v parameter in neutral kaons.

 We may envisage a connection between the cosmological processes responsible for the existence of matter--antimatter asymmetry and non-inertial effects by noting the similarities between different effects arising from QFTCS, particularly those concerning the thermodynamical phenomena stemming from the existence of event horizons in certain spacetimes~\cite{Hawking,Gibbons}, along with the observation that the temperature of the Universe is tied to particle creation due to its expansion~\cite{ParkerPRL,ParkerNature}. In what follows, we argue that investigations of mechanical phenomena such as the ones alluded above can play an important role in understanding thermodynamical effects, given this connection.

 Throughout this work we use natural units \(8\pi G = c = \hslash = k_{\textup{B}} = 1\), unless stated otherwise, and the \((-,+,+,+)\)~convention for the metric signature.

\section{\texorpdfstring{\(CP\)}{CP} violation in the \texorpdfstring{\(K\)}{K}-system}\label{CPvKsys}

 We start by briefly reviewing the basic ideas related to \(CP\)v in the kaon system~\cite{Greiner,Bigi}. In order to understand the unusual properties of neutral kaons, we first observe that, since both \(K^{0}\)~and~\(\,\overline{\!K}^{0}\) decay in two pions,
 \begin{equation}
  K^{0}\longleftrightarrow 2\pi\longleftrightarrow \,\overline{\!K}^{0},
  \label{cp2}
 \end{equation} 
 they are not independent particles with respect to the weak interaction. Charge conjugation~\(C\) and parity~\(P\) operations on kaon and pion states result in
 \begin{equation}
   CP\ket{K^{0}} = -\ket{\,\overline{\!K}^{0}}, \quad CP \ket{\pi\pi} = +\ket{\pi\pi}, \quad CP \ket{\pi\pi\pi} = -\ket{\pi\pi\pi}.
  \label{cp3}
 \end{equation}
 From these equations we see that the \(CP\)~operation does not leave the kaon states invariant, \(K^{0} \xrightarrow{CP} - \,\overline{\!K}^{0}\) and \(\,\overline{\!K}^{0} \xrightarrow{CP} - {K}^{0}\), therefore \(K^{0}\to 2\pi\) and \(\,\overline{\!K}^{0}\to 2\pi\) decays appear to be prohibited for \(CP\)-symmetric processes.

 If \(CP\)~symmetry is to be restored, we may consider, alternatively, linear combinations of theses states resulting in \(CP\)~eigenstates,
 \begin{equation}
   \ket{K_{1}}=\frac{1}{\sqrt{2}}\bigl(\ket{K^{0}}-\ket{\,\overline{\!K}^{0}}\bigr), \quad \ket{K_{2}}=\frac{1}{\sqrt{2}}\bigl(\ket{K^{0}}+\ket{\,\overline{\!K}^{0}}\bigr).
  \label{cp5}
 \end{equation}
 Using eq.~\eqref{cp3} one obtains
 \begin{equation}
  CP \ket{K_{1}} = +\ket{K_{1}}, \quad CP \ket{K_{2}} = -\ket{K_{2}}.
  \label{cp6}
 \end{equation}
 According to eq.~\eqref{cp3} and~\eqref{cp6}, \(K_{1}\) (\(K_{2}\)) can only decay into a state with eigenvalue \(CP=+1\) (\(CP=-1\)). Hence, we have only two possible decay channels: \(K_{1}\to 2 \pi\) and \(K_{2}\to 3 \pi\). Furthermore, a \(CP\)~projection operator can be defined,
 \begin{equation}
  \mathcal{P}_{+}=\frac{1}{2}(1 + CP), \quad \mathcal{P}_{-}=\frac{1}{2}(1 - CP),
  \label{cp6b}
 \end{equation}
 such that 
 \begin{equation}
  \mathcal{P}_{+}\ket{K_{1}} = \ket{K_{1}}, \quad
  \mathcal{P}_{-}\ket{K_{2}} = \ket{K_{2}}, \quad
  \mathcal{P}_{-}\ket{K_{1}} = \mathcal{P}_{+}\ket{K_{2}} = 0.
  \label{cp60}
 \end{equation}

\begin{figure}[t]
  \centering
  \includegraphics[width=.85\linewidth]{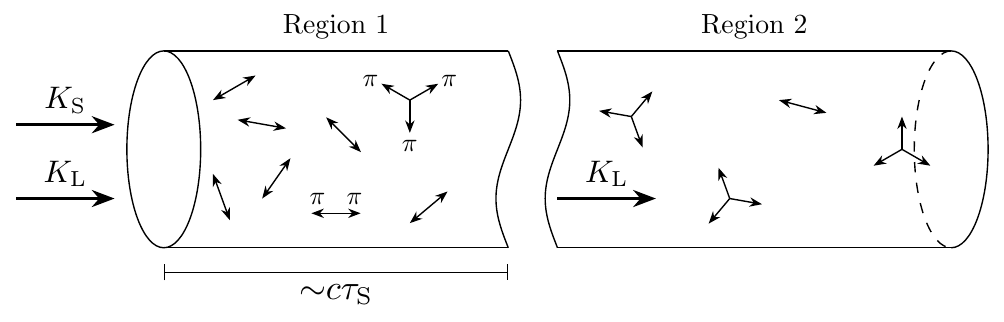}
  \caption{\label{cr-fig}Schematic view of the Cronin--Fitch experiment}
 \end{figure}

 To accommodate \(CP\)v in weak processes, the following mass eigenstates are introduced:
 \begin{equation}
  \ket{K_{\textup{S}}}=\frac{1}{\sqrt{1+\abs{q}^2}}\bigl(\ket{K^{0}} - q\ket{\,\overline{\!K}^{0}}\bigr), \quad \ket{K_{\textup{L}}}=\frac{1}{\sqrt{1+\abs{q}^2}}\bigl(\ket{K^{0}} + q\ket{\,\overline{\!K}^{0}}\bigr),
  \label{cp11}
 \end{equation}
 where the (complex) mixing parameter~\(q\) characterizes the strength of the \(CP\)v. There is a measurable difference in the lifetimes (with \(\tau_{\textup{S}} < \tau_{\textup{L}}\)) and masses (with \(m_{K_{\textup{S}}} < m_{K_{\textup{L}}}\)) of these eigenstates, justifying the ``S'' (referring to the \emph{short-lived} or \emph{small} state) and ``L'' (referring to the \emph{long-lived} or \emph{large} state) labels. We note that
 \begin{equation}
  \begin{aligned}
  \mathcal{P}_{+} \ket{K_{\textup{S}}} = \psi(+ q)\ket{K_{1}},& \quad
  \mathcal{P}_{-} \ket{K_{\textup{S}}} = \psi(- q)\ket{K_{2}},\\
  \mathcal{P}_{+} \ket{K_{\textup{L}}} = \psi(- q)\ket{K_{1}},& \quad
  \mathcal{P}_{-} \ket{K_{\textup{L}}} = \psi(+ q)\ket{K_{2}},
  \end{aligned}
  \label{cp11b}
 \end{equation}
 with
 \begin{equation}
  \psi(\pm q)=\frac{1}{\sqrt{2}}\frac{1\pm q}{\sqrt{1+\abs{q}^2}},
  \label{cp11c}
 \end{equation}
 which shows that the states in eq.~\eqref{cp11} are mixtures of \(CP = +1\) and \(CP = -1\) eigenstates. If \(q = 1\), the system is \(CP\)-symmetric and the projections in eq.~\eqref{cp11b} are reduced to those of eq.~\eqref{cp60}.
 The lifetimes of the states in question are~\cite{PDG}
 \begin{equation}
   \tau_{\textup{S}} = \SI{8.954(4)e-11}{\s}, \enspace \tau_{\textup{L}} = \SI{5.116(21)e-8}{\s},
  \label{cp7}
 \end{equation}
 with characteristic distances
 \begin{equation}
  c\tau_{\textup{S}} = \SI{2.684}{\cm}, \quad c\tau_{\textup{L}} = \SI{15.3}{\m}.
  \label{cp7b}
 \end{equation}
 In figure~\ref{cr-fig} we see a schematic representation of the Cronin--Fitch experiment~\cite{Cronin}, the first experimental result to provide clear evidence, from kaon decays, that \(CP\)~symmetry could be broken. An initially mixed beam of~\(K_{\textup{S}}\) and~\(K_{\textup{L}}\) is injected into a tube. The \(K_{\textup{S}}\)~particles traverse only their characteristic distance~\(c\tau_{\textup{S}}\) (region 1), decaying rapidly to~\(2 \pi\). The beam that emerges in region 2 is composed solely of \(K_{\textup{L}}\)~particles. Therefore, if \(CP\)~is a conserved symmetry (i.e., if~\(q=1\)) one should only observe the decay to~\(3 \pi\) in this region. \(CP\)v occurs because decays of type \(K_{\textup{L}} \to 2 \pi\), are also observed, implying that the weak eigenstates do not correspond to the \(CP\)~eigenstates.

 The \(CP\)v that occurs in the transition \(K \to 2 \pi\) can be studied by defining an observable~\(\eta\) that relates the the transition amplitudes~\(\mathcal{A}\) of the \(K_{\textup{S}}\) and \(K_{\textup{L}}\) species:
 \begin{equation}
  \eta_{00} \coloneqq \frac{\mathcal{A}(K_{\textup{L}} \to\pi^{0}\pi^{0})}{\mathcal{A}(K_{\textup{S}} \to\pi^{0}\pi^{0})},\quad \eta_{+-} \coloneqq \frac{\mathcal{A}(K_{\textup{L}} \to\pi^{+}\pi^{-})}{\mathcal{A}(K_{\textup{S}} \to\pi^{+}\pi^{-})}.
  \label{cp7d}
 \end{equation}
 Then, putting \(\eta = \eta_{00} = \eta_{+-}\) and using eq.~\eqref{cp11b} one obtains 
 \begin{equation}
  \eta = \frac{\braket{\pi\pi|\mathcal{P}_{+}| K_{\textup{L}}}}{\braket{\pi\pi|\mathcal{P}_{+}| K_{\textup{S}}}} = \frac{1-q}{1+q}.
  \label{cp7c}
 \end{equation}
 It must be noted that the experimental data suggests a difference between the absolute values of \(\eta_{00}\) and \(\eta_{+-}\)~\cite{PDG}:
 \begin{equation}
  \abs{\eta_{00}} = \num{2.220(11)e-3}, \quad \abs{\eta_{+-}} = \num{2.232(11)e-3}.
 \end{equation}
 In fact, in setting \(\eta_{+-} \equiv \epsilon + \epsilon^{\prime}\) and \(\eta_{+-} \equiv \epsilon - 2\epsilon^{\prime}\), the value of~\(\epsilon^{\prime}\) is nonzero but three orders of magnitude smaller then the value of~\(\epsilon\)~\cite{Bigi}. Alternatively, \(\abs{\eta}^{2}\)~is given in terms of the decay rates~\(\Gamma\) by
 \begin{equation}
  \abs{\eta}^2 = \frac{\Gamma(K_{\textup{L}}\to\pi\pi)}{\Gamma(K_{\textup{S}}\to\pi\pi)},
  \label{eta2}
 \end{equation}
 which we shall use in the subsequent analysis.

\section{Acceleration and \texorpdfstring{\(CP\)}{CP}-violating decays}\label{decayaccel}

 The Unruh effect (see~\cite{Crispino} for a review) states that accelerated observers perceive the inertial vacuum as a thermal bath of particles\footnote{More precisely, it is the observation that the vacuum state of a quantum field theory constructed with the time translation vector in Minkowski spacetime as a preferred direction corresponds, non-unitarily, to a Kubo--Martin--Schwinger (KMS) state in a quantum field theory for which the preferred direction vector is the Killing vector generating Lorentz boosts.}. It seems, then, that accelerated particles should have increased proper decay rates, as they would if immersed in a thermal bath. This is in fact the case, as was show for scalar fields by R. Müller in~\cite{Muller} and for spinorial fields by D. A. T. Vanzella and G. E. A. Matsas in~\cite{VanzellaShort,VanzellaLong}. These results also show a dependence of the rate of the increase on the mass of the decaying particle. Given the relationship between \(CP\)v and the kaon decays described above, we seek to investigate the impact that non-inertial effects have on \(CP\)v in the \(K\)-system with a model for the decay of accelerated kaons (related investigations, probing \(CP\)v in the lepton sector, can be found in~\cite{Ahluwalia,Blasone,Cozzella}). Such a model, based on the ones introduced by Müller, is presented below and used to evaluate the behavior of the decay rates and of the \(CP\)v parameter~\(\eta\), the latter of which being possible due the difference in the masses of \(K_{\textup{S}}\) and \(K_{\textup{L}}\).

\subsection{Model for accelerated decays}
 We consider the following interaction Lagrangian,
 \begin{equation}
  \mathcal{L}_{\textup{I}}(x) = G_{\Gamma}\Phi(x)\phi_{1}(x)\phi_{2}(x),
  \label{un1}
 \end{equation}
 where \(\Phi\)~is a scalar field of mass~\(M\), \(\phi_{1}\) and \(\phi_{2}\) are scalar fields of mass~\(m\) and \(G_{\Gamma}\)~is the coupling parameter of the interaction. Although kaons and pions are described by pseudoscalar fields, it is assumed that scalar fields provide approximate descriptions of their behavior. The decay rate may then be obtained from the decay amplitude, namely the transition amplitude for the process \(\Phi \to \phi_{1}\phi_{2}\), given, up to first order in~\(G_{\Gamma}\), by
 \begin{equation}
  \mathcal{A}(\mathbf{k}_{1},\mathbf{k}_{2}) = \bra{\mathbf{k}_{1},\mathbf{k}_{2}}\otimes\bra{0}S\ket{i}\otimes\ket{0} = G_{\Gamma}\int\iddn{4}{x}   \braket{0|\Phi(x)|i}\prod_{j=1}^{2}\braket{\mathbf{k}_{j}|\phi_{j}(x)|0},
  \label{un2}
 \end{equation}
 where the final state consists of two particles with momenta~\(\mathbf{k}_{1}\) and~\(\mathbf{k}_{2}\). The decay probability can be computed from the amplitude,
 \begin{equation}
  \begin{split}
   \mathcal{P} &= \int\iddn{3}{k_{1}}\iddn{3}{k_{2}}\abs{\mathcal{A}(\mathbf{k}_{1},\mathbf{k}_{2})}^{2} \\
   &= G_{\Gamma}^{2}\int\iddn{4}{x}\iddn{4}{x^{\prime}} \braket{0|\Phi(x)|i}\braket{i|\Phi(x^{\prime})|0}\prod_{j=1}^{2}\braket{0|\phi_{j}^{\dagger}(x)\int\iddn{3}{k_{j}}\ket{\mathbf{k}_{j}}\bra{\mathbf{k}_{j}}\phi_{j}(x^{\prime})|0}\\
   &= G_{\Gamma}^{2}\int\iddn{4}{x}\iddn{4}{x^{\prime}} f^{*}(x)f(x^{\prime})\prod_{j=1}^{2}\braket{0|\phi_{j}^{\dagger}(x)\phi_{j}(x^{\prime})|0},
  \end{split}
  \label{un3}
 \end{equation}
 where \(f(x)\)~is the mode associated to the initial state~\(\ket{i}\) and \(\braket{0|\phi_{j}^{\dagger}(x)\phi_{j}(x^{\prime})|0}\) is the Wightman function of~\(\phi_{j}\).

 Choosing a frame in Minkowski spacetime where \(\mathbf{x} = \mathbf{x}^{\prime}\) and assuming \(t>t^{\prime}\) (with \(x = (t,\mathbf{x})\) and \(x^{\prime} = (t^{\prime},\mathbf{x}^{\prime})\)), the Wightman function of a complex scalar field~\(\phi\) with mass~\(m\), for timelike separations of~\(x\) and~\(x^{\prime}\), reads
 \begin{multline}
  \braket{0|\phi^{\dagger}(x)\phi(x^{\prime})|0} = \frac{1}{(2\pi)^{3}}{\bra{0}}\int\iddn{3}{k}\frac{1}{\sqrt{2\omega(\mathbf{k})}}\Bigl(b(\mathbf{k})e^{ik\indices{_L}x\indices{^L}}+a^{\dagger}(\mathbf{k})e^{-ik\indices{_L}x\indices{^L}}\Bigr)\\
  \times\int\iddn{3}{k^{\prime}}\frac{1}{\sqrt{2\omega(\mathbf{k}^{\prime})}}\Bigl(a(\mathbf{k}^{\prime})e^{ik\indices*{^\prime_L}x\indices{^\prime^L}}+b^{\dagger}(\mathbf{k}^{\prime})e^{-ik\indices*{^\prime_L}x\indices{^\prime^L}}\Bigr){\ket{0}}.
 \end{multline}
 It is straightforward to show that the function above reduces to
 \begin{equation}
  \begin{split}
   \braket{0|\phi^{\dagger}(x)\phi(x^{\prime})|0} &= \frac{1}{(2\pi)^{3}}\int\iddn{3}{k}\frac{1}{2\omega(\mathbf{k})}e^{-i\omega(t-t^{\prime})}
   = \frac{1}{(2\pi)^{2}}\int_{\mathrlap{m}}^{\mathrlap{\infty}}\idd{\omega}\sqrt{\omega^{2}-m^{2}}e^{-i\omega(t-t^{\prime})}\\
   &= i\frac{m}{8\pi}\frac{H_{1}^{(2)}(m\Delta s)}{\Delta s},
  \end{split}
 \end{equation}
 where \(H_{1}^{(2)}\)~is a Hankel function of the second kind and \(\Delta s\)~is the spacetime interval of the timelike separated events~\(x\) and~\(x^{\prime}\).

 Assuming that \(f\)~is peaked over a trajectory~\(x(\tau)\) parameterized by its proper time~\(\tau\), i.e., over the trajectory of a particle, we can write it as
 \begin{equation}
  f(x) = h(\mathbf{x}(\tau))e^{-iM\tau}
  \label{un5}
 \end{equation}
 in the instantaneous rest frame. Assuming, furthermore, that the decay products do not deviate much from this trajectory, the decay probability can be written as
 \begin{equation}
  \begin{split}
   \mathcal{P} &= G_{\Gamma}^{2}\kappa\int_{-\infty}^{\infty}\int_{\mathrlap{-\infty}}^{\mathrlap{\infty}}\idd{\tau}\idd{\tau^{\prime}} e^{iM(\tau-\tau^{\prime})}\prod_{j=1}^{2}\braket{0|\phi_{j}^{\dagger}(t(\tau),\mathbf{x}(\tau))\phi_{j}(t^{\prime}(\tau^{\prime}),\mathbf{x}^{\prime}(\tau^{\prime}))|0}\\
   &= -\frac{G_{\Gamma}^{2}\kappa}{64\pi^{2}}m^{2}\int_{-\infty}^{\infty}\int_{\mathrlap{-\infty}}^{\mathrlap{\infty}}\idd{\tau}\idd{\tau^{\prime}}e^{iM(\tau-\tau^{\prime})}\frac{\bigl[H_{1}^{(2)}(m\Delta s)\bigr]^{2}}{\abs{\Delta s^{2}}},
  \end{split}
  \label{un6}
 \end{equation}
 where \(\kappa\)~is given by \(\kappa =\abs{\int\iddn{3}{x}h(\mathbf{x})}^2\). A change of variables of the form \(v \coloneqq \tau-\tau^{\prime}\) leads to
 \begin{equation}
  \mathcal{P} = -\frac{G_{\Gamma}^{2}\kappa}{64\pi^{2}}m^{2}\int_{-\infty}^{\infty}\int_{\mathrlap{-\infty}}^{\mathrlap{\infty}}\idd{v}\idd{\tau} e^{iMv}\frac{\bigl[H_{1}^{(2)}(m\Delta s)\bigr]^{2}}{\abs{\Delta s^{2}}}.
  \label{un7}
 \end{equation}
 Since the Wightman function depends only on the difference \(x-x^{\prime}\), the integral over~\(\tau\) is trivial (and infinite). We make use, thus, of the proper decay rate
 \begin{equation}
  \Gamma = -\frac{G_{\Gamma}^{2}\kappa}{64\pi^{2}}m^{2}\int_{\mathrlap{-\infty}}^{\mathrlap{\infty}}\idd{v} e^{iMv}\frac{\bigl[H_{1}^{(2)}(m\Delta s)\bigr]^{2}}{\abs{\Delta s^{2}}},
  \label{gammav}
 \end{equation}
 i.e., the decay probability by unit of proper time.

 A uniformly accelerated particle's trajectory may be parameterized in terms of the proper time as
 \begin{equation}
   t(\tau) = \frac{1}{a} \sinh(a\tau), \quad x(\tau) = \frac{1}{a} \cosh(a\tau), \quad y(\tau) = 0, \quad z(\tau) = 0,
  \label{un8}
 \end{equation}
 where \(a\)~is the magnitude of the proper acceleration. The squared spacetime interval~\(\Delta s^{2}\) at certain values~\(\tau\) and~\(\tau^{\prime}\) of the proper times is given by
 \begin{equation}
   \Delta s^{2} = -[t(\tau)-t^{\prime}(\tau^{\prime})]^{2} + [x(\tau)-x^{\prime}(\tau^{\prime})]^{2} = -\frac{4}{a^{2}}\sinh^{2}\Bigl(\frac{a}{2}\bigl(\tau-\tau^{\prime}\bigr)\Bigr).
  \label{un9}
 \end{equation}

 If we substitute \(\Delta s^{2}\) from \eqref{un9} into \eqref{gammav}, after introducing the variable \(u = a(\tau-\tau^{\prime})/2 = av/2\), we are led to
 \begin{equation}
  \Gamma = -\frac{G_{\Gamma}^{2}\kappa}{128\pi^{2}}m^{2}a\int_{\mathrlap{-\infty}}^{\mathrlap{\infty}}\idd{u}e^{i2Mu/a}\frac{\bigl\{H_{1}^{(2)}[2m\sinh(u)/a]\bigr\}^{2}}{\sinh^{2}(u)},
  \label{gamma}
 \end{equation}
 which is the decay rate for uniformly accelerated scalar particles.

\subsection{Results and Analysis}
 The experimental data relevant to the computation of eq.~\eqref{gamma} can be found in table~\ref{tabmass}, with the correspondences \(M = m_{K^{0}} \approx m_{K_{\textup{S}}} \approx m_{K_{\textup{L}}}\), \(m = m_{\pi^{0}} \approx m_{\pi^{\pm}}\)\footnote{Although the difference in the masses of \(\pi^{0}\) and \(\pi^{\pm}\) is not negligible, the results of the analysis are qualitatively the same for both cases.}. Besides the values in units of~\si{\mega\eV\!\per\lightspeed\squared}, table~\ref{tabmass} also includes values in units of the kaon mass in order to perform numerical computations. For the same reason, \(\Gamma\)~is given in units of \(G_{\Gamma}^{2}\kappa/(128\pi^{2})\).
 \begin{table}[t]
  \centering
  \begin{tabular*}{.75\columnwidth}{c@{\extracolsep{\fill}}S[table-format = 3.5(2)e+2]S[table-format = 1.6e+2]}
   \toprule &  \multicolumn{1}{c}{\si{\mega\eV\!\per\lightspeed\squared}} &  \multicolumn{1}{c}{\(m_{K^{0}}\)}\\
   \midrule
   \(m_{K^{0}}\) & 497.611(13) & 1\\
   \(m_{\pi^{0}}\) & 134.9770(5) & 0.271250\\
   \(m_{\pi^{\pm}}\) & 139.57061(24)& 0.280481\\
   \(m_{K_{\textup{L}}}-m_{K_{\textup{S}}}\) & 3.484(6)e-12 & 7.001e-15\\
   \bottomrule
  \end{tabular*}
  \caption{\label{tabmass}Values for the masses of the pions~\(\pi^{0}\), \(\pi^{\pm}\) and the neutral kaon~\(K^{0}\) and the mass difference between \(K_{\textup{L}}\)~and~\(K_{\textup{S}}\)~\cite{PDG}.}
 \end{table}

\begin{figure}[t]
  \centering
  \includegraphics[width=.6\linewidth]{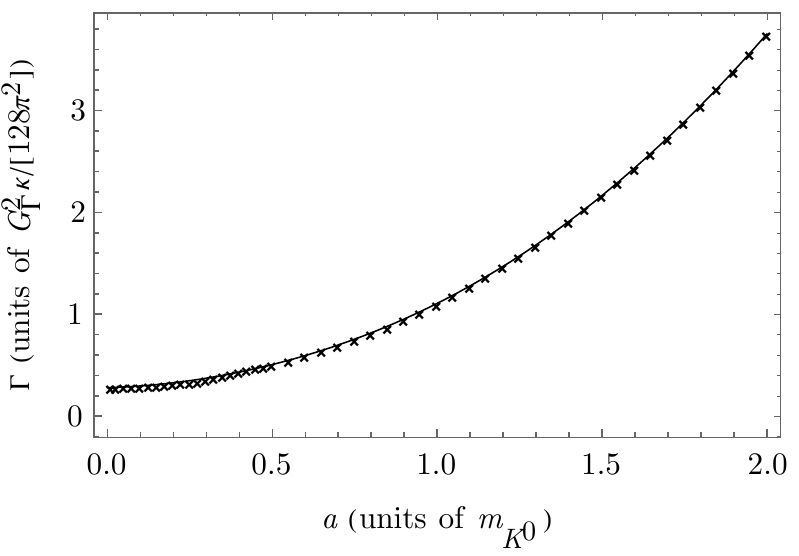}
  \caption{\label{plotgamma}Numerical results for the decay rate~\(\Gamma\) as a function of the acceleration~\(a\) and fitted curve.}
 \end{figure}

 The integral appearing in eq.~\eqref{gamma} presents several computational challenges which are discussed in appendix~\ref{HankelTreat}. The results of the aforementioned task, computed for varying values of the acceleration, are summarized in figure~\ref{plotgamma}, which also includes the graph of the function
 \begin{equation}
  \Gamma_{\textup{fit}}(a) = c_1 + c_2a^2 + c_3f(a) + c_4af(a) + c_5a^2f(a),
  \label{gamma-fit}
 \end{equation}
 where \(c_{1}=-7.00723\), \(c_{2}=0.329318\), \(c_{3}=7.28669\), \(c_{4}=0.0929348\), \(c_{5}=0.390045\) and
 \begin{equation}
  f(a)=\bigl(1-e^{-2\pi/a}\bigr)^{-1};
 \end{equation}
 the parameters \(c_{i}\), \(i=1, \ldots, 5\), were obtained by fitting the curve to the numerical data. The form of~\(\Gamma_{\textup{fit}}\) is based on the singular parts of the integral in eq.~\eqref{gamma} (given in eqs.~\eqref{singresults}).

 These results clearly indicate that the decay rate for this process grows as the acceleration increases, as expected from the results of previous analyses. It is also evident that this effect would only be detectable at extremely high energies, given that the acceleration is displayed in units of the kaon mass, which corresponds to accelerations of over~\SI{e32}{\meter\per\second\squared}. For comparison, the Texas Petawatt Laser, capable of accelerating electrons up to energies of~\(E=\SI{2}{\giga\eV}\) over distances of the order of \(\Delta x = \SI{1}{\cm}\)~\cite{TPLaser}, reaches an acceleration~\(a_{\textup{TPL}}\) given, at best, by
 \begin{equation}
  a_{\textup{TPL}} \approx \frac{E}{m_{e}\Delta x} \approx \SI[per-mode=symbol]{3.5e22}{\meter\per\second\squared}.
 \end{equation}
 There is also the problem of accelerating neutral particles, for which there is no obvious solution (see section~\ref{cosmoeff} for a discussion on why a cosmological setting may be better suited for the observation of effects of this kind).

 To investigate the behavior of~\(\eta\) it is sufficient, having eq.~\eqref{eta2} in mind, to compute~\(\Gamma\) for the two values of~\(M\) corresponding to the \(K_{\textup{S}}\) and \(K_{\textup{L}}\) masses and take their ratio. Since the difference between these values is very small, we take \(m_{K_{\textup{S}}} \approx m_{K^{0}}\) and \(m_{K_{\textup{L}}} \approx m_{K^{0}} + (m_{K_{\textup{L}}} - m_{K_{\textup{S}}})\). A small inconvenience for this computation is the fact that \(G_{\Gamma}(K_{\textup{S}}\to\pi\pi)\) and \(G_{\Gamma}(K_{\textup{L}}\to\pi\pi)\) must be different to account for the considerable difference in the magnitudes of \(\Gamma(K_{\textup{S}}\to\pi\pi)\) and \(\Gamma(K_{\textup{L}}\to\pi\pi)\). This may be remedied by considering a rescaled \(CP\)v parameter~\(\eta^{\prime}\), given by
 \begin{equation}
  \eta = \beta\eta^{\prime}, \qquad \beta = \frac{G_{\Gamma}(K_{\textup{L}}\to\pi\pi)}{G_{\Gamma}(K_{\textup{S}}\to\pi\pi)}.
 \end{equation}

\begin{figure}[t]
  \centering
  \includegraphics[width=.6\linewidth]{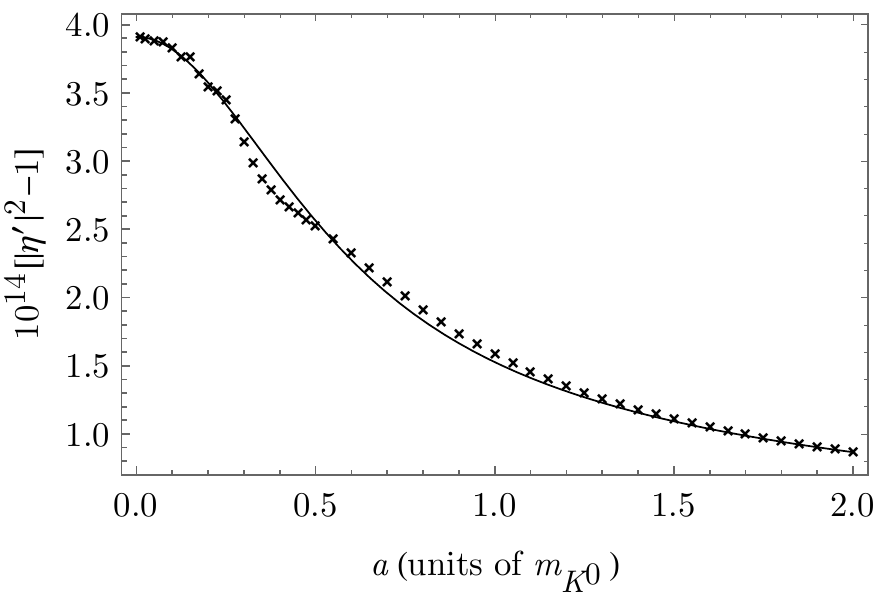}
  \caption{\label{ploteta}Numerical results for~\(\abs{\eta^{\prime}}^{2}-1\) as a function of the acceleration~\(a\) and plot of an approximation obtained from singular integrals.}
 \end{figure}

 The numerical results for~\(\abs{\eta^{\prime}}^{2}-1\), as well as a plot of an approximation of it obtained from the singular parts of eq.~\eqref{gamma} (see appendix~\ref{HankelTreat} for more details), can be found in figure~\ref{ploteta}. While they point to a variation of~\(\abs{\eta}^{2}\) with increasing acceleration, which implies that \(CP\)v is sensitive to non-inertial effects, it is a decrease---which indicates that the contribution of these kaon decays to \(CP\)v is smaller at very high acceleration scales---of the order of~\(\num{e-13}\abs{\eta}^{2}\) (currently impossible to detect, even at extremely high accelerations, given that the experimental uncertainty for~\(\eta\) is of at least~0.4\%). The tiny fluctuations of the numerical data at small acceleration scales can be attributed to errors in the numerical calculation of the integral for the decay rate.

\begin{figure*}[t]
  \includegraphics[width=\linewidth]{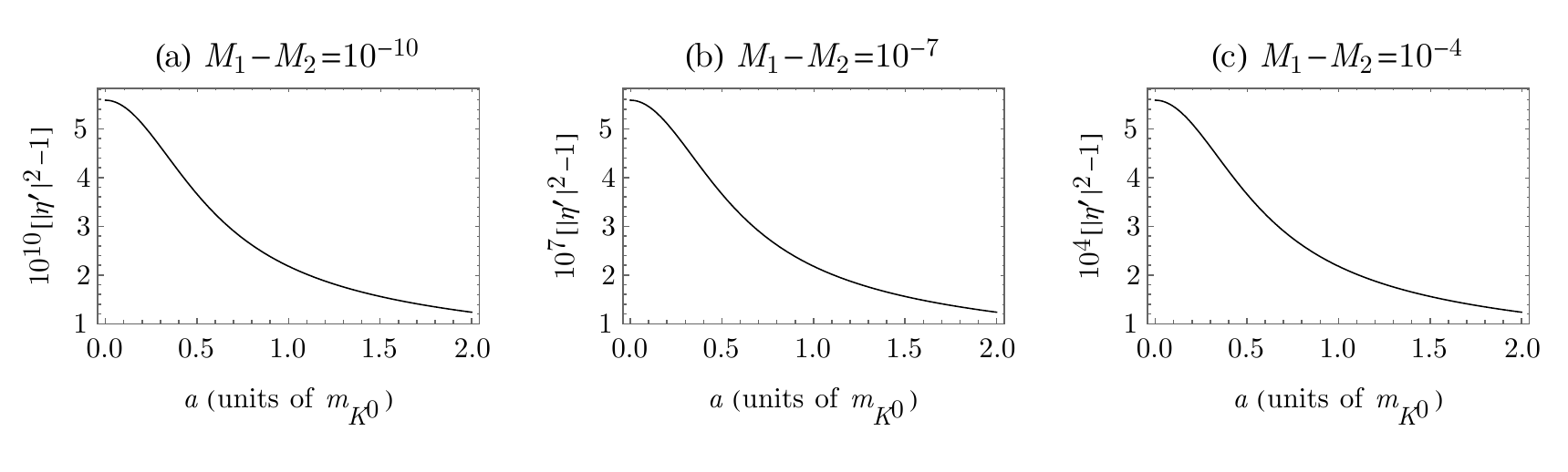}
  \caption[etadif]{\label{plotetadif}Plot of~\(\abs{\eta^{\prime}}^{2} - 1\) for different values of the mass difference: \begin{enumerate*}[label=(\alph*)]
   \item \(\num{e-10}m_{K^{0}}\),
   \item \(\num{e-7}m_{K^{0}}\) and
   \item \(\num{e-4}m_{K^{0}}\).
  \end{enumerate*}}
 \end{figure*}

 Further investigation of the behavior of~\(\eta^{\prime}\) using the singular parts of the decay rate integrals leads to an interesting relationship between the amplitude of the variation of~\(\eta^{\prime}\) and the value of the relative mass difference between the two species modeled by the field~\(\Phi\). As an example of the kind of physics this would entail, consider, in place of \(K_{\textup{S}}\) and \(K_{\textup{L}}\), two species \(\Phi^{(1)}\) and \(\Phi^{(2)}\) with masses (in units of \(m_{K^{0}}\)) \(M_{1} = 1 + (M_{1}-M_{2}) > 1\) and \(M_{2} = 1\), respectively. Using the singular integral approximation mentioned above, the computation of
 \begin{equation}
  \eta^{\prime} = \frac{G_{\Gamma}\bigl(\Phi^{(2)}\to\phi_{1}^{(2)}\phi_{2}^{(2)}\bigr)}{G_{\Gamma}\bigl(\Phi^{(1)}\to\phi_{1}^{(1)}\phi_{2}^{(1)}\bigr)}\frac{\Gamma\bigl(\Phi^{(1)}\to\phi_{1}^{(1)}\phi_{2}^{(1)}\bigr)}{\Gamma\bigl(\Phi^{(2)}\to\phi_{1}^{(2)}\phi_{2}^{(2)}\bigr)},
 \end{equation}
 where \(\phi_{i}^{(j)}\) are scalar fields of mass~\(m = m_{\pi^{0}}/m_{K^{0}}\), for different values of \(M_{1}-M_{2}\) yields the results show in figure~\ref{plotetadif}. They indicate a roughly linear relation between the mass difference and the amplitude of the fluctuation of~\(\abs{\eta^{\prime}}^{2} - 1\), from which one may infer that the scale of the effect is given by the relative mass difference~\((M_{1}-M_{2})/M_{2}\) (or, in the case of the kaon decays, \((m_{K_{\textup{L}}} - m_{K_{\textup{S}}})/m_{K^{0}}\)). It seems reasonable to infer that for a species with~\(M_{1}<M_{2}\) there should be an increase in the magnitude of~\(\eta\), still proportional to the mass difference~\(M_{2}-M_{1}\). This is the case for \(K_{\textup{S}} \to 3\pi\) processes, which are also \(CP\)-violating~\cite{PDG}, though a model for decays of this kind would be slightly different. The nature of the results on the \(K_{\textup{S}} \to 3\pi\) channel is, nevertheless, expected to be the same as for the \(K_{\textup{L}} \to 2\pi\) channel (see~\cite{Muller} for a comparison of computations using the method discussed above for different models).

\section{Cosmological effects}\label{cosmoeff}

 An intimate relationship between non-inertial and cosmological effects exists in QFTCS: the presence of an event horizon is correlated to the existence of non-Hilbert--Schmidt Bogoliubov transformations, i.e., QFT constructions that are not unitarily equivalent, which manifest themselves as correspondences between vacuum and thermal (KMS)~states (see~\cite{WaldQFT} for an overview). This is the case for the aforementioned Unruh effect, the well known Hawking effect for black holes~\cite{Hawking} and the Gibbons--Hawking effect in de~Sitter spacetime~\cite{Gibbons}, all of which present Killing horizons. A similar effect is that of particle creation in expanding universes, first described by Parker in~\cite{ParkerPRL,ParkerNature}, which predicts a thermal signature for particles created by the expansion of a Friedmann--Lemaître--Robertson--Walker (FLRW) universe. This section seeks to draw parallels between these effects and to encourage further research on the impact of phenomena stemming from QFTCS on other areas of physics by discussing the extension of the predictions of section~\ref{decayaccel} to cosmological settings.

 A direct connection between the Unruh effect and the Gibbons--Hawking effect is established by noting that both are due to the presence of a Killing horizon. The Killing fields generating these horizons are the Lorentz boost generators for the Unruh effect and boost-like generators for the Gibbons--Hawking effect (in fact, these vector fields can be lifted to Lorentz boost generators on a 5-dimensional Minkowski spacetime where the de~Sitter spacetime can be embedded). The surface gravity~\(\kappa\) of these horizons is related to the temperature of the KMS~state by
 \begin{equation}
  T = \frac{\kappa}{2\pi},
 \end{equation}
 with \(\kappa = a\) for the Unruh effect and \(\kappa = \sqrt{\Lambda/3}\) for the Gibbons--Hawking effect, where \(\Lambda\)~is the cosmological constant. One may then wonder if \(CP\)-violating processes in de~Sitter spacetime would behave like those of accelerated particles: would an observer also measure the difference in the behaviour of the observables discussed above when compared to measurements by an inertial observer in Minkowski spacetime? This seems reasonable (so long as \(a = \sqrt{\Lambda/3}\)), given that both the accelerated observer in Minkowski spacetime and the observer in de~Sitter spacetime perceive KMS~states of the same temperature. A computation of decay rates of particles in a thermal bath is discussed in~\cite{Matsas,MatsasTest}, as is the relationship between it and the results obtained in~\cite{VanzellaShort,VanzellaLong}, which firmly establishes the complementarity of the non-inertial and thermal effects.
 
 With respect to particle creation in expanding universes (i.e., in FLRW~universes), the connection is more tenuous: though there exists a kind of horizon in solutions in this class, it is not (in general) a Killing horizon but a particle horizon. Nevertheless, the prediction of QFTCS is that the expansion leads to particle creation at a temperature
 \begin{equation}
  T \approx \frac{1}{\sqrt{2\pi}}\frac{S_{1}}{S(t)},
 \end{equation}
 where \(S(t)\)~is the scale factor and \(S_{1}\)~is some lower bound for the scale factor (see~\cite{ParkerNature} for more details). This allows one to draw the same conclusion as above (so long as \(a \approx \sqrt{2\pi}S_{1}/S\)). More realistic cosmological models, such as the \(\Lambda\)CDM model, present both particle and event horizons (the latter being closely related to the de~Sitter event horizon, given the presence of the cosmological constant), leading to a combination of effects, though the Gibbons--Hawking component is very small for our Universe, since \(\Lambda = \num{7.15(19)e-121}\)~\cite{Planck2018} (compare to \(a_{\textup{TPL}}^{2} \approx \num{e-60}\)).

 In the context of the thermal history of the Universe, kaons freeze-out near the QCD crossover, at a temperature \(T_{\textup{QCD}} \approx \SI{155}{\mega\eV\!\per\boltzmannk}\)~\cite{HotQCD}, the moment at which the \(CP\)-violating processes discussed above are expected to contribute to the observed matter--antimatter asymmetry. Inverting the expression for the Unruh temperature allows the determination of the acceleration scale corresponding to the crossover temperature, \(a_{\textup{QCD}} = 2\pi T_{\textup{QCD}} \approx 2 m_{K^{0}}\). Figure~\ref{ploteta} then allows us to infer that, at the crossover, \(\eta\)~would be near to its lowest value, implying a lower amplitude of \(CP\)v when compared to current values and diminished matter--antimatter asymmetry.

\section{Conclusions}

 The model introduced above for the \(K \to 2\pi\) decays allows for the computation of the decay rate over accelerated trajectories, corroborating the results of ref.~\cite{Muller}. The subsequent analysis of the behaviour of the \(CP\)v parameter~\(\eta\) leads to the conclusion that its squared magnitude~\(\abs{\eta}^{2}\) decreases very slightly with increasing acceleration (around three parts in \num{e14} for an acceleration \(a \approx 2m_{K_{0}} \approx \SI{4e32}{\meter\per\second\squared}\)). We have shown that the amplitude of the decrease is proportional to the mass difference between the weak eigenstates (for the kaons, \((m_{K_{\textup{L}}}-m_{K_{\textup{L}}})/m_{K^{0}} \approx \num{e-14}\)). 

 The discussion on the relationship between the Unruh effect, the Gibbons--Hawking effect and the phenomenon of particle creation in accelerating universes led to the argument that, given the complementarity between non-inertial and thermal effects, a similar conclusion may hold for fields in de~Sitter and FLRW universes (though the proper computations still need to be executed to fully justify this affirmation): \(CP\)v contributions of this nature are smaller at higher temperatures in these spacetimes.

 Further research on the connection between the results presented in this work and those on the thermal dependence of \(CP\)v in the quark sector is in progress. A decrease in the effectiveness of \(CP\)-violating processes in this sector with increasing temperature has been shown in~\cite{Brauner}, but the relation between it and our result is not immediately clear.

\appendix
\section{Numerical Treatment of the Decay Rate}\label{HankelTreat}
 In this appendix, the numerical treatment of the decay rate, eq.~\eqref{gamma}, is presented. The integral appearing in this expression is
 \begin{equation}
  \mathcal{I} = \int_{\mathrlap{-\infty}}^{\mathrlap{\infty}}\idd{u} e^{i2Mu/a}\frac{\bigl\{H_{1}^{(2)}[2m\sinh(u)/a]\bigr\}^{2}}{\sinh^{2}(u)}.
  \label{apen1}
 \end{equation}
 It is difficult to tackle this expression analytically, so the numerical approach is favored. Problems in the implementation of the numerical methods appear due to singularities in this expression, which means that the singular parts of the integral must be separated into an integral~\(\mathcal{I}_{0}\) and treated analytically, so that \(\mathcal{I}\)~can be computed as
 \begin{equation}
  \mathcal{I} = \underbrace{(\mathcal{I} - \mathcal{I}_{0})}_{\mathclap{\substack{\text{treated}\\\text{numerically}}}}\mkern8mu + \mkern8mu \underbrace{\mathcal{I}_{0}}_{\mathclap{\substack{\text{treated}\\\text{analytically}}}}.
 \end{equation}

 The singular part of~\(\mathcal{I}\) can be determined using the power series of the Bessel functions of first and second kind, \(J_{1}\)~and~\(Y_{1}\), given in equations~10.2.2 and~10.8.1 of ref.~\cite{DLMF}, up to third order to determine the singular parts of the Hankel function of second kind~\(H_{1}^{(2)}\):
 \begin{equation}
  \begin{split}
   H_{1}^{(2)}(z) &= J_{1}(z) - iY_{1}(z)\\
   &=\frac{2i}{\pi z} - \frac{3iz^{3}}{32\pi} + \biggl(z - \frac{z^{3}}{8}\biggr)\biggl\{\frac{1}{2} - \frac{i}{\pi}\biggl[\ln\Bigl(\frac{z}{2}\Bigr) - \frac{1}{2} + \gamma\biggr]\biggr\} + \order{z^{4}},
  \end{split}
  \label{apen2}
 \end{equation} 
 where \(\gamma\)~is the Euler--Mascheroni constant. Since this function appears squared in the expression for the decay rate and also involves a division by a second order polynomial on the argument of function, \(2m \sinh(u)/a\), the following expression is computed:
 \begin{equation}
  \begin{split}
   \frac{\bigl[H_{1}^{(2)}(2mz/a)\bigr]^{2}}{z^{2}} &= 2\biggl[\frac{i}{\pi}+\frac{2}{\pi^{2}}\biggl(\gamma-\frac{1}{2}\biggr)\biggr]\frac{1}{z^{2}} - \frac{a^{2}}{\pi^{2}m^{2}}\frac{1}{z^{4}} + \frac{4}{\pi^{2}}\frac{1}{z^{2}}\ln\Bigl(\frac{m}{a}z\Bigr)\\
   &\pheq + \frac{2m^{2}}{\pi^{2}a^{2}}(1-4\gamma-2i\pi)\ln\Bigl(\frac{m}{a}z\Bigr) - \frac{4m^{2}}{\pi^{2}a^{2}}\ln^{2}\Bigl(\frac{m}{a}z\Bigr) + \order{1}.
  \end{split}\label{hankelseries}
 \end{equation}
 The integral~\(\mathcal{I}_{0}\) of the singular parts is then given by
 \begin{equation}
  \mathcal{I}_{0} = \sum_{j=1}^{5}\mathcal{I}_{j},
 \end{equation}
 where
 \begin{subequations}
  \begin{align}
   \mathcal{I}_{1} &\coloneqq -\frac{a^{2}}{\pi^{2}m^{2}}\int_{\mathrlap{-\infty}}^{\mathrlap{\infty}}\idd{u}\frac{e^{i2Mu/a}}{\sinh^{4}(u)},\\
   \mathcal{I}_{2} &\coloneqq 2\biggl[\frac{i}{\pi}+\frac{2}{\pi^{2}}\biggl(\gamma-\frac{1}{2}\biggr)\biggr]\int_{\mathrlap{-\infty}}^{\mathrlap{\infty}}\idd{u}\frac{e^{i2Mu/a}}{\sinh^{2}(u)},\\
   \mathcal{I}_{3} &\coloneqq \frac{4}{\pi^{2}}\int_{\mathrlap{-\infty}}^{\mathrlap{\infty}}\idd{u}\frac{e^{i2Mu/a}}{\sinh^{2}(u)}\ln\Bigl(\frac{m}{a}\sinh(u)\Bigr),\\
   \mathcal{I}_{4} &\coloneqq \frac{2m^{2}}{\pi^{2}a^{2}}(1-4\gamma-2i\pi)\int_{\mathrlap{-\infty}}^{\mathrlap{\infty}}\idd{u}e^{i2Mu/a}\ln\Bigl(\frac{m}{a}\sinh(u)\Bigr),\\
   \mathcal{I}_{5} &\coloneqq -\frac{4m^{2}}{\pi^{2}a^{2}}\int_{\mathrlap{-\infty}}^{\mathrlap{\infty}}\idd{u}e^{i2Mu/a}\ln^{2}\Bigl(\frac{m}{a}\sinh(u)\Bigr).
  \end{align}
 \end{subequations}

 The integrals~\(\mathcal{I}_{1}\) and~\(\mathcal{I}_{2}\) can be calculated using the residue theorem,
 \begin{align}
  \mathcal{I}_{1} &\propto \frac{8\pi M}{3a^{3}}\frac{a^{2}+M^{2}}{1-e^{-2\pi M/a}},\\
  \mathcal{I}_{2} &\propto -\frac{4\pi M}{a}\frac{1}{1-e^{-2\pi M/a}}.
  \label{apen4}
 \end{align}
 Of note is that the contributions of~\(\mathcal{I}_{1}\) and~\(\mathcal{I}_{2}\) to the decay rate correspond to the decay rates of the processes~\(\Psi \to \psi_{1} \psi_{2}\) and~\(\Psi \to \psi_{1}\) respectively, where~\(\Psi\) is a massive scalar field and~\(\psi_{1}\) and \(\psi_{2}\)~are massless scalar fields, since the Wightman function of a massless scalar field over an accelerated trajectory is proportional to \(1/\sinh^{2}(u)\) (see~\cite{Muller} for an explicit calculation involving massless fields).
 The integrals~\(\mathcal{I}_{4}\) and~\(\mathcal{I}_{5}\) can be approximated by integrals~\(\mathcal{I}_{4}^{\prime}\) and~\(\mathcal{I}_{5}^{\prime}\) with similar singularity structures,
 \begin{align}
  \begin{split}
   \mathcal{I}_{4}^{\prime} &= \frac{2m^{2}}{\pi^{2}a^{2}}(1-4\gamma-2i\pi)\int_{\mathrlap{-\infty}}^{\mathrlap{\infty}}\idd{u}e^{-2M\abs{u}/a}\ln\Bigl(\frac{m}{a}\abs{u}\Bigr)\\
   &=-\frac{2m^{2}}{\pi^{2}aM}(1-4\gamma-2i\pi)\biggl[\gamma+\ln\biggl(\frac{2M}{m}\biggr)\biggr],
  \end{split}
  \label{I4'}\\
  \begin{split}
   \mathcal{I}_{5}^{\prime} &= -\frac{4m^{2}}{\pi^{2}a^{2}}\int_{\mathrlap{-\infty}}^{\mathrlap{\infty}}\idd{u}e^{-2M\abs{u}/a}\Bigl[\ln^2\Bigl(\frac{m}{a}\abs{u}\Bigr)+2i\pi\theta(-u)\ln\Bigl(\frac{m}{a}\abs{u}\Bigr)\Bigr]\\
   &= -\frac{4m^{2}}{\pi^{2}aM}\Biggl\{\frac{\pi^{2}}{6}+\biggl[\gamma+\ln\biggl(\frac{2M}{m}\biggr)\biggr]^{2}-i\pi\biggl[\gamma+\ln\biggl(\frac{2M}{m}\biggr)\biggr]\Biggr\}.
  \end{split}
  \label{I5'}
 \end{align}
 
 Solving these integrals makes use of the parity of the integrands and equations~4.331\nobreakdash--1 and~4.335\nobreakdash--1 of ref.~\cite{Gradshtein}.

 Computing~\(\mathcal{I}_{3}\) can be achieved using the following identity for the logarithm,
 \begin{equation}\label{logid}
  \ln(z) = \lim_{w \to 0}\frac{z^{w}-1}{w}.
 \end{equation}
 Then,
 \begin{equation}
  \mathcal{I}_{3} = \frac{4}{\pi^{2}}\int_{\mathrlap{-\infty}}^{\mathrlap{\infty}}\idd{u}e^{i2Mu/a}\ln\Bigl(\frac{m}{a}\sinh(u)\Bigr)\biggl[\frac{1}{u^{2}}-\frac{1}{3} + \order{u}\biggr].
 \end{equation}
 This implies that \(\mathcal{I}_{3}\)~can be approximated by two other integrals, as in eqs.~\eqref{I4'} and~\eqref{I5'},
 \begin{align}
  \mathcal{I}_{3}^{\prime} &= \frac{4}{\pi^{2}}\int_{\mathrlap{-\infty}}^{\mathrlap{\infty}}\idd{u}\frac{e^{i2Mu/a}}{u^{2}}\ln\Bigl(\frac{m}{a}u\Bigr),\\
  \mathcal{I}_{3}^{\prime\prime} &= -\frac{4}{3\pi^{2}}\int_{\mathrlap{-\infty}}^{\mathrlap{\infty}}\idd{u}e^{-2M\abs{u}/a}\ln\Bigl(\frac{m}{a}u\Bigr).
 \end{align}
 The solution for the integral~\(\mathcal{I}_{3}^{\prime\prime}\) follows directly from eq.~\eqref{I4'},
 \begin{equation}
  \mathcal{I}_{3}^{\prime\prime} = \frac{4a}{3\pi^{2}M}\biggl[\gamma+\ln\biggl(\frac{2M}{m}\biggr)\biggr],
 \end{equation}
 while the one for~\(\mathcal{I}_{3}^{\prime}\) requires the use of eq.~\eqref{logid}:
 \begin{equation}
  \begin{split}
   \mathcal{I}_{3}^{\prime} &= \frac{4}{\pi^{2}}\int_{\mathrlap{-\infty}}^{\mathrlap{\infty}}\idd{u}\frac{e^{i2Mu/a}}{u^{2}}\ln\Bigl(\frac{m}{a}u\Bigr) \propto \lim_{w \to 0}\int_{\mathrlap{-\infty}}^{\mathrlap{\infty}}\idd{u}\frac{e^{i2Mu/a}}{u^{2}}\frac{(mu/a)^{w}-1}{w}\\
   &\propto \lim_{w \to 0}\frac{(-i)^{w-2}}{w}\Bigl(\frac{m}{a}\Bigr)^{w}\int_{\mathrlap{-\infty}}^{\mathrlap{\infty}}\idd{u}e^{i2Mu/a}(iu)^{^{w-2}} - \lim_{w \to 0}\frac{(-i)^{-2}}{w}\int_{\mathrlap{-\infty}}^{\mathrlap{\infty}}\idd{u}\frac{e^{i2Mu/a}}{(iu)^{2}}.
  \end{split}
 \end{equation}
 The solutions to the integrals appearing above can be found in equations~3.382\nobreakdash--6 and 3.382\nobreakdash--7 of ref.~\cite{Gradshtein}. Therefore,
 \begin{equation}
  \begin{split}
   \mathcal{I}_{3}^{\prime} &\propto -\lim_{w \to 0}\frac{1}{w}\Bigl(-\frac{im}{a}\Bigr)^{w}\biggl(\frac{2M}{a}\biggr)^{1-w}\frac{2\pi}{\Gamma(2-w)} + \lim_{w \to 0}\frac{1}{w}\frac{2M}{a}\frac{2\pi}{\Gamma(2)}\\
   &\propto -\frac{4\pi M}{a}\lim_{w \to 0}\frac{1}{w}\Bigl(-\frac{im}{2M}\Bigr)^{w}[1+(1-\gamma)w] + \lim_{w \to 0}\frac{1}{w}\frac{4\pi M}{a}\\
   &\propto -\frac{4\pi M}{a}\lim_{w \to 0}\biggl\{\frac{1}{w}\Bigl[\Bigl(-\frac{im}{2M}\Bigr)^{w}-1\Bigr] +\Bigl(\frac{-im}{2M}\Bigr)^{w}(1-\gamma)\biggr\}\\
   &\propto -\frac{4\pi M}{a}\Bigl[\ln\Bigl(-\frac{im}{2M}\Bigr)+1-\gamma\Bigr] = -\frac{4\pi M}{a}\biggl[\ln\Bigl(\frac{m}{2M}\Bigr)-\frac{i\pi}{2}+1-\gamma\biggr],
  \end{split}
 \end{equation}
 where it was used that~\(1/\Gamma(2-w) = 1+(1-\gamma)w\).

 The strategy introduced at the beginning of this appendix can now be executed, but the singular integral~\(\mathcal{I}_{0}\) must be substituted by
 \begin{equation}
  \mathcal{I}_{0}^{\prime} = \sum_{j=1}^{2}\mathcal{I}_{j} + \sum_{j=3}^{5}\mathcal{I}_{j}^{\prime} + \mathcal{I}_{3}^{\prime\prime},
 \end{equation}
 where
 \begin{subequations}\label{singresults}
  \begin{align}
   \mathcal{I}_{1} &= -\frac{8 M}{3\pi am^{2}}\frac{a^{2}+M^{2}}{1-e^{-2\pi M/a}},\\
    \mathcal{I}_{2} &= -\frac{8\pi M}{a}\biggl[\frac{i}{\pi}+\frac{2}{\pi^{2}}\biggl(\gamma-\frac{1}{2}\biggr)\biggr]\frac{1}{1-e^{-2\pi M/a}},\\
   \mathcal{I}_{3}^{\prime} &= -\frac{16 M}{\pi a}\biggl[\ln\Bigl(\frac{m}{2M}\Bigr)-\frac{i\pi}{2}+1-\gamma\biggr],\\
    \mathcal{I}_{4}^{\prime} &= -\frac{2m^{2}}{\pi^{2}aM}(1-4\gamma-2i\pi)\biggl[\gamma+\ln\biggl(\frac{2M}{m}\biggr)\biggr],\\
    \mathcal{I}_{5}^{\prime} &= -\frac{4m^{2}}{\pi^{2}aM}\Biggl\{\frac{\pi^{2}}{6}+\biggl[\gamma+\ln\biggl(\frac{2M}{m}\biggr)\biggr]^{2}-i\pi\biggl[\gamma+\ln\biggl(\frac{2M}{m}\biggr)\biggr]\Biggr\},\\
   \mathcal{I}_{3}^{\prime\prime} &= \frac{4a}{3\pi^{2}M}\biggl[\gamma+\ln\biggl(\frac{2M}{m}\biggr)\biggr],
  \end{align}
 \end{subequations}
 so that \(\mathcal{I}\) can be computed as
 \begin{equation}
  \mathcal{I} = \underbrace{(\mathcal{I} - \mathcal{I}_{0}^{\prime})}_{\mathclap{\substack{\text{treated}\\\text{numerically}}}}\mkern8mu + \mkern8mu \underbrace{\mathcal{I}_{0}^{\prime}}_{\mathclap{\substack{\text{treated}\\\text{analytically}}}}.
 \end{equation}
 The integral~\((\mathcal{I} - \mathcal{I}_{0}^{\prime})\) is highly oscillatory with a rapidly decaying integrand, requiring special techniques to be used in its computation. The dominant contribution to~\(\Gamma\) comes from~\(\mathcal{I}_{0}^{\prime}\), especially for high values of~\(a\) (see eq.~\eqref{hankelseries}). The results also present very small imaginary parts (at their largest, 2~orders of magnitude smaller than their real parts) that can be attributed to numerical errors.

%%%%%%%%%%%%%%%%%%%%%%%%%%%%%%%%%%%%%%%%%%%%%%%%%%%%%%%%%%%%%%%%%%%%%%%

\section*{Acknowledgments}
The authors would like to thank George E. A. Matsas for enlightening comments. The research was supported by Conselho Nacional de Desenvolvimento Científico e Tecnológico (CNPq) under the grant 130658/2017-0

\bibliography{Bib.bib}

\end{document}